\documentclass[12pt]{article}
\usepackage[left=2.5cm,top=2.50cm,right=2.5cm,bottom=2.50cm]{geometry}
\usepackage{mathrsfs}
\usepackage{amsmath,amssymb,latexsym,color,cancel,graphicx,bbm,colortbl}
\usepackage[english]{babel}
\usepackage[latin1]{inputenc}
\usepackage{ragged2e}
\usepackage{cite}
\begin{document}
\date{}

\title{Exact Solution of the Relativistic Dunkl Oscillator in $(2+1)$ Dimensions}
\author{R. D. Mota$^{a,}$, D. Ojeda-Guill\'en$^{b,c}$\footnote{{\it E-mail address:} dojedag@ipn.mx},\\ M. Salazar-Ram{\'i}rez$^{b}$ and V. D. Granados$^{c}$} \maketitle

\begin{minipage}{0.9\textwidth}
\small $^{a}$ Escuela Superior de Ingenier{\'i}a Mec\'anica y El\'ectrica, Unidad Culhuac\'an,
Instituto Polit\'ecnico Nacional, Av. Santa Ana No. 1000, Col. San
Francisco Culhuac\'an, Del. Coyoac\'an, C.P. 04430, Ciudad de M\'exico, Mexico.\\

\small $^{b}$ Escuela Superior de C\'omputo, Instituto Polit\'ecnico Nacional,
Av. Juan de Dios B\'atiz esq. Av. Miguel Oth\'on de Mendiz\'abal, Col. Lindavista,
Del. Gustavo A. Madero, C.P. 07738, Ciudad de M\'exico, Mexico.\\

\small $^{c}$ Escuela Superior de F{\'i}sica y Matem\'aticas,
Instituto Polit\'ecnico Nacional, Ed. 9, Unidad Profesional Adolfo L\'opez Mateos, Del. Gustavo A. Madero, C.P. 07738, Ciudad de M\'exico, Mexico.\\

\end{minipage}

\begin{abstract}
In this paper we study the $(2+1)$-dimensional Dirac-Dunkl oscillator coupled to an external magnetic field. Our Hamiltonian is obtained from the standard Dirac oscillator coupled to an external magnetic field by changing the partial derivatives by the Dunkl derivatives. We solve the Dunkl-Klein-Gordon-type equations in polar coordinates in a closed form. The angular part eigenfunctions are given in terms of the Jacobi-Dunkl polynomials and the radial functions in terms of the Laguerre functions. Also, we compute the energy spectrum of this problem and show that, in the non-relativistic limit, it properly reduces to the Hamiltonian of the two dimensional harmonic oscillator.

\end{abstract}

PACS: 02.30.Ik, 02.30.Jr, 03.65.Ge, 03.65.Pm\\
Keywords: Dirac equation, Dirac-Moshinsky oscillator, Dunkl derivative

\section{Introduction}

The relativistic version of the Dirac oscillator was not so direct to obtain in Quantum Mechanics. To achieve this, Ito \emph{et al.} \cite{Ito} and Cook \cite{Cook} added the linear term $-imc\omega\beta{\mathbf{\alpha}}\cdot \mathbf{r}$ to the relativistic momentum $\textbf{p}$ of the free-particle Dirac equation. Later, this problem was called ``Dirac oscillator" by Moshinsky and Szczepaniak who constructed its explicit solutions, showed that its symmetry Lie algebra is $so(4)\oplus so(3,1)$ and found its generators \cite{Mos,Mos2}. A peculiar characteristic of this problem is that in the non-relativistic limit, it reduces to the harmonic oscillator plus a spin-orbit coupling term. The Dirac-Moshinsky oscillator is a crucial problem in Relativistic Quantum Mechanics and has been extensively studied and applied in many branches of physics, as can be seen in Refs. \cite{Delgado,Sadurni,Ferkous,Villalba,Benitez,Moreno,deLima}. One of the most important applications of the Dirac oscillator is its connection with Quantum Optics \cite{Bermudez,Sadurni,Mandal} through the Jaynes-Cummings and Anti-Jaynes-Cummings models \cite{Jaynes}.

The Dunkl operators $D_{x_i}$ are combinations of differential and difference operators, associated to a finite reflection group $\mathcal{G}$, and were introduced by Yang in the context of the deformed oscillator \cite{YANG}. Dunkl reintroduced these operators to study polynomials in several variables with reflection symmetries related to finite reflection groups \cite{DUNKL}. The Dunkl Laplacian, obtained from the Dunkl derivative, is a combination of the classical Laplacian in $\mathbb{R}^n$ with some difference terms, such that the resulting operator is only invariant under $\mathcal{G}$ and not under the whole orthogonal group \cite{DUNKL,XuY}. The Dunkl operators are closely related to certain representations of degenerate affine Hecke algebras \cite{Chere,Opdam}, and its commutative algebra has been used to study integrable models of quantum mechanics, as the Calogero-Sutherland-Moser models \cite{HIK,KAK,LAP}.

Recently, the quantum mechanical isotropic oscillator and Coulomb problems involving the Dunkl derivatives have been shown to be superintegrable systems, and are closely related to the $-1$ orthogonal polynomials of the Bannai-Ito scheme \cite{GEN1,GEN2,GEN3,GEN4}. In these works, the authors also obtained the symmetry of each problem (called Schwinger-Dunkl algebra) and the exact solutions of the Schr\"odinger equation in terms of Jacobi, Laguerre and Hermite polynomials. In the same sense, in Refs. \cite{NOS1,NOS2} we solved the two-dimensional Dunkl-oscillator and the Dunkl-Coulomb problems algebraically by using the $su(1,1)$ Lie algebra and the theory of unitary irreducible representations.

In particular, in Refs. \cite{GEN1,GEN4} the harmonic oscillator and the Coulomb problems have been studied by changing the standard derivatives $\frac{\partial}{\partial_{x_i}}$ by the Dunkl derivatives $D_{x_i}\equiv\frac{\partial}{\partial x_i}+\frac{\mu_{x_i}}{x}(1-R_{x_i})$, $x_i=x, y$. This is why these systems are referred  as Dunkl-oscillator and the Dunkl-Coulomb problems, respectively. In the present paper we extend this idea to the relativistic regime, by studying the Dirac oscillator coupled to an external magnetic field, with Dunkl derivatives. We refer this problem as the Dirac-Dunkl oscillator. The aim of the present work is to study the exact analytical solutions of the $(2+1)$-dimensional Dirac-Dunkl oscillator coupled to an external magnetic field. For this problem, we obtain in closed form the eigenfunctions and the energy spectrum.

This work is organized as follows. In Section $2$, we introduce the Dirac-Dunkl oscillator in $2+1$ dimensions coupled to an external magnetic field. We decouple the equations for the upper and lower wave functions and introduce polar coordinates to separate variables. Along this work we use the results of Ref. \cite{GEN1} to solve the angular part. In Section $3$, we obtain the radial states and energy spectrum for the upper and the lower components when the condition $\omega>\omega_c/2$ is satisfied.  In Section $4$, we study the cases $\omega=\omega_c/2$ and $\omega_c/2>\omega$ with an analogous procedure to the one we used in Section $3$ to obtain its exact solutions. In Section $5$, we compute the non-relativistic limit of our problem to show that our Hamiltonian can be properly reduced to the non-relativistic harmonic oscillator in two dimensions. Finally, we give some concluding remarks.

\section{Dirac-Dunkl Hamiltonian of the Oscillator Coupled to an External Magnetic Field}

The time-independent Dirac equation for the Dirac-Moshinsky oscillator is given by the Hamiltonian \cite{Mos}
\begin{equation}
H_D\Psi=\left[c \mathbf{\alpha} \cdot\left(\mathbf{p}-im\omega \mathbf{r}\beta\right)+mc^2\beta\right]\Psi=E\Psi.
\end{equation}
If we consider the presence of an external uniform magnetic field $\mathbf{B}$, this Hamiltonian is modified as
\begin{equation}
H_D=c \mathbf{\alpha} \cdot\left[\left(\mathbf{p}-\frac{e\mathbf{A}}{c}\right)-im\omega \mathbf{r}\beta\right]+mc^2\beta,\label{hamc}
\end{equation}
where $\omega$ is the oscillator frequency, $\mathbf{r}$ is the position vector of the oscillator and $\mathbf{A}$ is the vector potential
in the symmetric gauge
\begin{equation}
\mathbf{A}=\left(-\frac{By}{2},\frac{Bx}{2}\right).
\end{equation}
Now, we restrict our analysis to $2+1$ dimensions in which the Dirac matrices are set in the standard way in terms of Pauli spin matrices
$\alpha_1=\sigma_1, \alpha_2=\sigma_2, \beta=\sigma_3$. Thus, the Hamiltonian of equation (\ref{hamc}) takes the form
\begin{equation}
H_D=\begin{pmatrix}
mc^2 &-i\hbar c \left(\frac{\partial}{\partial x}-i\frac{\partial}{\partial  y}\right)+imc\tilde{\omega}(x-iy){} \\
-i\hbar c \left(\frac{\partial}{\partial x}+i\frac{\partial}{\partial y}\right)-imc\tilde{\omega} (x+iy) & -mc^2 \end{pmatrix},\label{matriz}
\end{equation}
where $\tilde{\omega}=\omega-\frac{\omega_c}{2}$ and $\omega_c=\frac{|e|B}{m}$ is the cyclotron frequency. If we change in this Hamiltonian the
partial derivatives $\frac{\partial}{\partial x}$ and $\frac{\partial}{\partial y}$ by the Dunkl derivatives
\begin{equation}
D_x\equiv\frac{\partial}{\partial x}+\frac{\mu_x}{x}(1-R_x), \quad\quad D_y\equiv\frac{\partial}{\partial y}+\frac{\mu_y}{y}(1-R_y),
\end{equation}
we obtain what we call the Dirac-Dunkl Hamiltonian $H_{DD}$. The constants in the Dunkl derivative are restricted to be $\mu_x\geq -\frac{1}{2}$ and $\mu_y\geq -\frac{1}{2}$ \cite{GEN1}. Here, $R_{x}$ and $R_y$ are the reflection operators in the $x-$ and $y-$ coordinates respectively, it is to say,
$R_xf(x,y)=f(-x,y)$ and $R_yf(x,y)=f(x,-y)$.

The purpose of the present paper is to solve the eigenvalues problem $H_{DD}\psi=E\psi$, where $\psi=(\psi_1(\mathbf{r}), \psi_2(\mathbf{r}))^T$.
Using the properties of the reflection operators $R_xR_y=R_yR_x$, $R_{x}^2=1$,  $\frac{\partial}{\partial x}R_{x}=-R_{x}\frac{\partial}{\partial x}$, $R_{x}x=-xR_{x}$, and similar expressions for the $y-$ coordinate, we decouple the Dirac-Dunkl
differential equation for the upper and lower spinor components $\psi_1$ and $\psi_2$. Explicitly, we obtain
\begin{eqnarray}
\left(-\frac{1}{2}{\hbar}^2 c^2 \nabla_{D}^2+\hbar m\tilde{\omega} c^2\left(i(xD_y-yD_x)-\mu_xR_x-\mu_yR_y-1\right)+\frac{1}{2}m^2\tilde{\omega}^2c^2(x^2+y^2) \right)\psi_1=\mathcal{E}\psi_1,\label{klein1}\\
\left(-\frac{1}{2}{\hbar}^2 c^2 \nabla_{D}^2+\hbar m\tilde{\omega} c^2\left(i(xD_y-yD_x)+\mu_xR_x+\mu_yR_y+1\right)+\frac{1}{2}m^2\tilde{\omega}^2c^2(x^2+y^2) \right)\psi_2=\mathcal{E}\psi_2,\label{klein2}
\end{eqnarray}
where we have defined $\mathcal{E}=\frac{E^2-m^2c^4}{2}$. In these expressions, the Dunkl Laplacian $\nabla_{D}^2$ is defined as
\begin{equation}
\nabla_D^2=\frac{\partial^2}{\partial x^2}+\frac{\partial^2}{\partial y^2}+2\frac{\mu_x}{x}\frac{\partial}{\partial x}+2\frac{\mu_y}{y}\frac{\partial}{\partial y}+\frac{\mu_x}{x^2}(1-R_x)+\frac{\mu_y}{y^2}(1-R_y).
\end{equation}
By using polar coordinates, this operator takes the form
\begin{equation}
-\frac{1}{2}\nabla_D^2=-\frac{1}{2}\left(\frac{\partial^2}{\partial \rho^2}+\frac{1}{\rho}\frac{\partial}{\partial \rho}\right)-\frac{1}{\rho}(\mu_x+\mu_y)\frac{\partial}{\partial \rho}+\frac{1}{\rho^2}B_\phi,
\end{equation}
where $B_\phi$ is defined in equation (\ref{be}) of Appendix. We emphasize that the angular operator $\mathcal{J}\equiv i(xD_y-yD_x)$ is such
$\mathcal{J}^2=2B_\phi+2\mu_x\mu_y(1-R_xR_y)$. This allows us to write the Dunkl-Laplacian operator as
\begin{equation}
-\frac{1}{2}\nabla_D^2=-\frac{1}{2}\left(\frac{\partial^2}{\partial \rho^2}+\frac{1}{\rho}\frac{\partial}{\partial \rho}\right)-\frac{1}{\rho}(\mu_x+\mu_y)\frac{\partial}{\partial \rho}+\frac{1}{\rho^2}\left(\frac{\mathcal{J}^2-2\mu_x\mu_y(1-R_xR_y)}{2}\right).
\end{equation}
With this equation and the definition of the operator $\mathcal{J}$ we can rewrite the Dunkl-Klein-Gordon equations (\ref{klein1}) and (\ref{klein2}) as follows
\begin{eqnarray}
-\frac{1}{2}\frac{\partial^2\psi_1}{\partial \rho^2}-\frac{1}{\rho}\left(\frac{1}{2}+\mu_x+\mu_y\right)\frac{\partial \psi_1}{\partial \rho}+\frac{1}{2}\left(\frac{\mathcal{J}^2-2\mu_x\mu_y(1-R_xR_y)}{\rho^2}\right)\psi_1+\frac{m\tilde{\omega}}{\hbar}\mathcal{J}\psi_1\nonumber\\
-\frac{m\tilde{\omega}}{\hbar}(1+\mu_xR_x+\mu_yR_y)\psi_1+\frac{1}{2}\frac{m^2\tilde{\omega}^2}{\hbar^2}\rho^2 \psi_1=\tilde{\mathcal{E}}\psi_1,\label{eq1}\\
-\frac{1}{2}\frac{\partial^2\psi_2}{\partial \rho^2}-\frac{1}{\rho}\left(\frac{1}{2}+\mu_x+\mu_y\right)\frac{\partial \psi_2}{\partial \rho}+\frac{1}{2}\left(\frac{\mathcal{J}^2-2\mu_x\mu_y(1-R_xR_y)}{\rho^2}\right)\psi_2+\frac{m\tilde{\omega}}{\hbar}\mathcal{J}\psi_2\nonumber\\
+\frac{m\tilde{\omega}}{\hbar}(1+\mu_xR_x+\mu_yR_y)\psi_2+\frac{1}{2}\frac{m^2\tilde{\omega}^2}{\hbar^2}\rho^2 \psi_2=\tilde{\mathcal{E}}\psi_2,\label{eq2}
\end{eqnarray}
were $\tilde{\mathcal{E}}\equiv \frac{\mathcal{E}}{\hbar^2 c^2}=\frac{E^2-m^2c^4}{2\hbar^2 c^2}$. Now, the main problem is to solve the above differential equations. Similarly as it occur in the case of the non-relativistic Dunkl oscillator and the Dunkl-Coulomb problems \cite{GEN1,GEN4}, as we will show in Section 3, the Dunkl-Klein-Gordon equations (\ref{eq1}) and (\ref{eq2}) of the relativistic oscillator can be solved by separable variables in polar coordinates. This is achieved due to the eigenvalues and eigenfunctions of the Dunkl angular operator $\mathcal{J}$ that have been found previously in Ref. \cite{GEN1}. In the Appendix we have given a summary of these results.

It has been shown that the Dunkl derivatives are anti-Hermitian
\begin{equation}
\langle f|D_x|g\rangle=-\langle g|D_x|f\rangle, \hspace{5ex}\langle f|D_y|g\rangle=-\langle g|D_y|f\rangle,
\end{equation}
with respect to the scalar product \cite{GEN1}
\begin{eqnarray}
\langle f|g\rangle\equiv \int_{-\infty}^{\infty}\int_{-\infty}^{\infty}f^*(x,y)g(x,y)|x|^{2\mu_x}|y|^{2\mu_y}dxdy\hspace{9.5ex}\nonumber\\
\hspace{15ex}=\int_{0}^{\infty}\int_{0}^{2\pi}f^*(\rho,\theta)g(\rho,\theta)|\rho \cos\theta|^{2\mu_x}|\rho \sin\theta|^{2\mu_y}\rho d\rho d\theta.\label{sp}
\end{eqnarray}
Therefore, the operators $iD_x$ and $iD_y$ are Hermitian for all functions $f$ and $g$ belonging to the $\mathcal{L}^2$-space associated with the above scalar product. The Hamiltonian (\ref{matriz}) with Dunkl derivatives leads to the equations
\begin{eqnarray}
-i\hbar c \left(D_x-iD_y\right)|\psi_2\rangle+imc\tilde\omega (x-iy)|\psi_2\rangle=(E-mc^2)|\psi_1\rangle\label{DD1},\\
-i\hbar c \left(D_x+iD_y\right)|\psi_2\rangle-imc\tilde\omega (x+iy)|\psi_1\rangle=(E-mc^2)|\psi_2\rangle\label{DD2}.
\end{eqnarray}
By multiplying on the left equation (\ref{DD1}) by $\langle\psi_1|$ and equation (\ref{DD2}) by $\langle\psi_2|$, respectively, and using that
$iD_x$ and $iD_y$ are Hermitian, we show that
\begin{equation}
(E-mc^2)\langle\psi_1|\psi_1\rangle=(E+mc^2)\langle\psi_2|\psi_2\rangle.
\end{equation}
This equality and the normalization condition that must satisfy the upper and lower Dirac spinor components
\begin{equation}
\langle\psi_1|\psi_1\rangle+\langle\psi_2|\psi_2\rangle=1,
\end{equation}
lead to obtain the following results
\begin{equation}
\langle\psi_1|\psi_1\rangle=\sqrt{\frac{E+mc^2}{2E}},\hspace{8ex}\langle\psi_2|\psi_2\rangle=\sqrt{\frac{E-mc^2}{2E}}.
\end{equation}
Thus, we conclude that the energy-normalized upper and lower spinor components for the Dirac-Dunkl oscillator are
\begin{equation}
|\psi_1'\rangle\equiv \sqrt{\frac{E+mc^2}{2E}}|\psi_1\rangle,\hspace{8ex}|\psi_2'\rangle\equiv \sqrt{\frac{E-mc^2}{2E}}|\psi_2\rangle.
\end{equation}

\section{Exact Analytical Solutions for the case $\omega>\omega_c/2$}

This Section is dedicated to obtain the analytical solutions of equations (\ref{eq1}) and (\ref{eq2}) of the upper and lower spinors $\psi_1$ and  $\psi_2$ if we assume explicitly $\omega>\omega_c/2$. Our results are based on the fact that the operator $R_xR_y$ commutes with the operator $\mathcal{J}$ and with the Dunkl-Klein-Gordon Hamiltonian-type operators (left-hand side of equations (\ref{klein1}) and (\ref{klein2})).
By using the results of Appendix, we find that there are two possibilities, namely $R_x=R_y$ and $R_x=-R_y$. In what follows, we shall study these cases separately and report the upper and lower radial spinor solutions which are finite at the coordinates origin.

\subsection{Case $R_x=R_y$}
In this case $\epsilon\equiv s_xs_y=1$, with $s_x=s_y=1$ or $s_x=s_y=-1$, being $s_x$ and $s_y$ the eigenvalues of $R_x$ and $R_y$, respectively. Also, $\mathcal{J}F_+=\lambda_+F_+$, where $F_+=\Phi_n^{++}(\phi)\pm i\Phi_n^{--}(\phi)$, $\lambda_+=\pm2\sqrt{n(n+\mu_x+\mu_y)}$, $n\in {\mathbb{N}}$ and $\Phi_n^{++}$ and $\Phi_n^{--}$ are given by equations (\ref{masmas}) and  (\ref{menmen}) of the Appendix. The additional term to $\mathcal{J}^2$ in the centrifugal potential vanishes. \\

For $s_x=s_y=1$, we set $\psi_1^{++}=\psi_1^{++}(\rho)F_+(\phi)$. The radial solutions $\psi_1^{++}(\rho)$ are obtained by solving equation (\ref{eq1}) with the operator $\mathcal{J}$ replaced by its eigenvalue $\lambda_+$. Thus, we obtain
\begin{equation}
\psi_1^{++}(\rho)=\rho^{\sqrt{\lambda_+^2+(\mu_x+\mu_y)^2}-\mu_x-\mu_y} e^{-\frac{1}{2}\frac{m\tilde{\omega}}{\hbar}\rho^2}L_{k}^{\sqrt{\lambda_+^2+(\mu_x+\mu_y)^2}}\left(m\tilde{\omega}\rho^2/\hbar\right),\label{f1}\\
\end{equation}
where we have defined $\frac{1}{2m\tilde{\omega}}(\tilde{\mathcal{E}}\hbar+m\tilde{\omega}(\mu_x+\mu_y-\lambda_+)-m\tilde{\omega}\sqrt{\lambda_+^2+(\mu_x+\mu_y)^2})=k$, $k=0, 1, 2, ...$. Since  $\tilde{\mathcal{E}}_k=\frac{E_k^2-m^2c^4}{2\hbar^2 c^2}$, we obtain that the energy spectrum for this spinor component is
\begin{equation}
E_k^{++}=\pm mc^2\sqrt{1+\frac{4\tilde{\omega}\hbar}{mc^2}k+\frac{2\tilde{\omega}\hbar}{mc^2}\sqrt{\lambda_+^2+(\mu_x+\mu_y)^2}-\frac{2\tilde{\omega}\hbar}{mc^2}(\mu_x+\mu_y-\lambda_+)}.
\end{equation}
Similarly, we find from equation (\ref{eq2}) that the radial part of $\psi_2^{++}=\psi_2^{++}(\rho)F_+(\phi)$ is given by
\begin{equation}
\psi_2^{++}(\rho)=\rho^{\sqrt{\lambda_+^2+(\mu_x+\mu_y)^2}-\mu_x-\mu_y} e^{-\frac{1}{2}\frac{m\tilde{\omega}}{\hbar}\rho^2}L_{k'}^{\sqrt{\lambda_+^2+(\mu_x+\mu_y)^2}}\left(m\tilde{\omega}\rho^2/\hbar\right),\label{f2}
\end{equation}
and the energy spectrum
\begin{equation}
E_{k'}^{++}=\pm mc^2\sqrt{1+\frac{4\tilde{\omega}\hbar}{mc^2}k'+\frac{2\tilde{\omega}\hbar}{mc^2}\sqrt{\lambda_+^2+(\mu_x+\mu_y)^2}+\frac{2\tilde{\omega}\hbar}{mc^2}(\mu_x+\mu_y+\lambda_++2)}.
\end{equation}
Since both components $\psi_1^{++}$ and $\psi_2^{++}$ belong to the same Dirac-Dunkl spinor, they correspond to the same energy, $E_{k}^{++}=E_{k'}^{++}$.
This requirement sets up a constraint between the quantum numbers $k'$ and $k$. In fact
\begin{equation}
E_{k}^{++}=E_{k'}^{++}\Rightarrow k'=k-\mu_x-\mu_y-1,\label{cond1}
\end{equation}
with $k,k' \in {\mathbb{N}}$.\\

For the case $s_x=s_y=-1$, also the additional term to $\mathcal{J}^2$ in the centrifugal potential vanishes. The upper wave function is $\psi_1^{--}=\psi_1^{--}(\rho)F_+(\phi)$, with $\mathcal{J}^2F_+(\phi)=\lambda_+F_+(\phi)$. Therefore, the upper part for the radial spinor is found by
solving equation (\ref{eq1}). We obtain
\begin{equation}
\psi_1^{--}(\rho)=\rho^{\sqrt{\lambda_+^2+(\mu_x+\mu_y)^2}-\mu_x-\mu_y} e^{-\frac{1}{2}\frac{m\tilde{\omega}}{\hbar}\rho^2}L_{\ell}^{\sqrt{\lambda_+^2+(\mu_x+\mu_y)^2}}\left(m\tilde{\omega}\rho^2/\hbar\right),\label{f1--}\\
\end{equation}
with its respective energy spectrum given by
\begin{equation}
E_\ell^{--}=\pm mc^2\sqrt{1+\frac{4\tilde{\omega}\hbar}{mc^2}\ell+\frac{2\tilde{\omega}\hbar}{mc^2}\sqrt{\lambda_+^2+(\mu_x+\mu_y)^2}+\frac{2\tilde{\omega}\hbar}{mc^2}(\mu_x+\mu_y+\lambda_+)}.
\end{equation}
From equation (\ref{eq2}) we obtain for the lower radial spinor component $\psi_2^{--}=\psi_2^{--}(\rho)F_+(\phi)$
\begin{equation}
\psi_2^{--}(\rho)=\rho^{\sqrt{\lambda_+^2+(\mu_x+\mu_y)^2}-\mu_x-\mu_y} e^{-\frac{1}{2}\frac{m\tilde{\omega}}{\hbar}\rho^2}L_{\ell'}^{\sqrt{\lambda_+^2+(\mu_x+\mu_y)^2}}\left(m\tilde{\omega}\rho^2/\hbar\right),\label{f2--}
\end{equation}
and
\begin{equation}
E_{\ell'}^{--}=\pm mc^2\sqrt{1+\frac{4\tilde{\omega}\hbar}{mc^2}\ell'+\frac{\tilde{2\omega}\hbar}{mc^2}\sqrt{\lambda_+^2+(\mu_x+\mu_y)^2}-\frac{2\tilde{\omega}\hbar}{mc^2}(\mu_x+\mu_y-\lambda_+-2)}.
\end{equation}
The condition that both functions $\psi_1^{--}$ and $\psi_2^{--}$ belong to the same energy leads to the restriction $E_{\ell'}^{--}=E_{\ell}^{--}$ which is satisfied if
\begin{equation}\label{cond2}
\ell'=\ell+\mu_x+\mu_y-1,
\end{equation}
with $\ell,\ell' \in {\mathbb{N}}$.\\

\subsection{Case $R_x=-R_y$}
Here, we shall consider $\epsilon =s_xs_y=-1$ and there are two possibilities: $s_x=1$ and $s_y=-1$ or $s_x=-1$ and $s_y=1$. As the previous case, according to the results of the Appendix, we have $\mathcal{J}F_-=\lambda_-F_-$, where $F_-=\Phi_n^{-+}(\phi)\mp i\Phi_n^{+-}(\phi)$ and $\lambda_-=\pm2\sqrt{(n+\mu_x)(n+\mu_y)}$, $n\in \{\frac{1}{2},\frac{3}{2},...\}$.

For $s_x=1$ and $s_y=-1$, the centrifugal factor is $\frac{1}{2}(\mathcal{J}^2-4\mu_x\mu_y)$. Further, the fifth term in equation (\ref{eq1}) for $\psi_1$ is now given by $-\frac{m\tilde{\omega}}{\hbar}(1+\mu_x-\mu_y)$. Under these considerations, we find the radial part of the eigenfunctions $\psi_1^{+-}=\psi_1^{+-}(\rho)F_-(\phi)$ from equation (\ref{eq1}). They are given by
\begin{equation}
\psi_1^{+-}(\rho)=\rho^{\sqrt{\lambda_-^2+(\mu_x-\mu_y)^2}-\mu_x-\mu_y} e^{-\frac{1}{2}\frac{m\tilde{\omega}}{\hbar}\rho^2}L_{\tilde{k}}^{\sqrt{\lambda_-^2+(\mu_x-\mu_y)^2}}\left(m\tilde{\omega}\rho^2/\hbar\right),\label{f3}\\
\end{equation}
where it was  defined
$\frac{1}{2m\tilde{\omega}}(\tilde{\mathcal{E}}\hbar+m\tilde{\omega}(\mu_x-\mu_y-\lambda_-)-m\tilde{\omega}\sqrt{\lambda_-^2+(\mu_x-\mu_y)^2})=\tilde{k}$, $\tilde k=0, 1, 2, ...$. Since  $\tilde{\mathcal{E}}_{\tilde k}=\frac{E_{\tilde k}^2-m^2c^4}{2\hbar^2 c^2}$, we obtain the energy spectrum
\begin{equation}
E_{\tilde k}^{+-}=\pm mc^2\sqrt{1+\frac{4\tilde{\omega}\hbar}{mc^2}{\tilde k}+\frac{2\tilde{\omega}\hbar}{mc^2}\sqrt{\lambda_-^2+(\mu_x-\mu_y)^2}-\frac{2\tilde{\omega}\hbar}{mc^2}(\mu_x-\mu_y-\lambda_-)}.
\end{equation}
Similarly, for $\psi_2^{+-}=\psi_2^{+-}(\rho)F_-(\phi)$ from equation (\ref{eq2}) we obtain
\begin{equation}
\psi_2^{+-}(\rho)=\rho^{\sqrt{\lambda_-^2+(\mu_x-\mu_y)^2}-\mu_x-\mu_y} e^{-\frac{1}{2}\frac{m\tilde{\omega}}{\hbar}\rho^2}L_{\tilde{k}'}^{\sqrt{\lambda_-^2+(\mu_x-\mu_y)^2}}\left(m\tilde{\omega}\rho^2/\hbar\right),\label{f3+-}\\
\end{equation}
and the spectrum
\begin{equation}
E_{\tilde k'}^{+-}=\pm mc^2\sqrt{1+\frac{4\tilde{\omega}\hbar}{mc^2}{\tilde k'}+\frac{2\tilde{\omega}\hbar}{mc^2}\sqrt{\lambda_-^2+(\mu_x-\mu_y)^2}+\frac{2\tilde{\omega}\hbar}{mc^2}(\mu_x-\mu_y+\lambda_-+2)}.
\end{equation}
Also, the condition $E^{+-}_{\tilde k}=E^{+-}_{\tilde k'}$ imposes a restriction between the quantum numbers ${\tilde k}$ and ${\tilde k}'$ given by
\begin{equation}
\tilde{k}'=\tilde{k}-\mu_x+\mu_y-1,\label{cond3}
\end{equation}
with ${\tilde k},{\tilde k}' \in {\mathbb{N}}$.\\

Finally, we consider the second possibility of this case $s_x=-1$, $s_y=1$. Also, the centrifugal factor is $\frac{1}{2}(\mathcal{J}^2-4\mu_x\mu_y)$ and the fifth term in equation (\ref{eq1}) for $\psi_1$ is given by $-\frac{m\tilde{\omega}}{\hbar}(1-\mu_x+\mu_y)$. Thus, from equation (\ref{eq1}) we find  that the radial part of the eigenfunctions $\psi_1^{-+}=\psi_1^{-+}(\rho)F_-(\phi)$ are given by
\begin{equation}
\psi_1^{-+}(\rho)=\rho^{\sqrt{\lambda_-^2+(\mu_x-\mu_y)^2}-\mu_x-\mu_y} e^{-\frac{1}{2}\frac{m\tilde{\omega}}{\hbar}\rho^2}L_{\bar{k}}^{\sqrt{\lambda_-^2+(\mu_x-\mu_y)^2}}\left(m\tilde{\omega}\rho^2/\hbar\right),\label{f3-+2}\\
\end{equation}
and the energy spectrum
\begin{equation}
E_{\bar{k}}^{-+}=\pm mc^2\sqrt{1+\frac{4\tilde{\omega}\hbar}{mc^2}{\bar{k}}+\frac{2\tilde{\omega}\hbar}{mc^2}\sqrt{\lambda_-^2+(\mu_x-\mu_y)^2}+\frac{2\tilde{\omega}\hbar}{mc^2}(\mu_x-\mu_y+\lambda_-)}.
\end{equation}
From equation (\ref{eq2}) we find the radial part of the spinor component $\psi_2^{-+}=\psi_2^{-+}(\rho)F_-(\phi)$
\begin{equation}
\psi_2^{-+}(\rho)=\rho^{\sqrt{\lambda_-^2+(\mu_x-\mu_y)^2}-\mu_x-\mu_y} e^{-\frac{1}{2}\frac{m\tilde{\omega}}{\hbar}\rho^2}L_{\bar{k}'}^{\sqrt{\lambda_-^2+(\mu_x-\mu_y)^2}}\left(m\tilde{\omega}\rho^2/\hbar\right),\label{f3-+}\\
\end{equation}
and
\begin{equation}
E_{\bar{k}'}^{-+}=\pm mc^2\sqrt{1+\frac{4\tilde{\omega}\hbar}{mc^2}{\bar{k}'}+\frac{2\tilde{\omega}\hbar}{mc^2}\sqrt{\lambda_-^2+(\mu_x-\mu_y)^2}-\frac{2\tilde{\omega}\hbar}{mc^2}(\mu_x-\mu_y-\lambda_--2)}.
\end{equation}
The condition $E^{-+}_{\bar{k}}=E^{-+}_{\bar{k}'}$ imposes
\begin{equation}
\bar{k}'=\bar{k}+\mu_x-\mu_y-1,\label{cond4}
\end{equation}
with $\bar{k},\bar{k}' \in {\mathbb{N}}$.\\

Therefore, because of conditions (\ref{cond1}), (\ref{cond2}), (\ref{cond3}) and (\ref{cond4}), in order to have square-integrable Laguerre functions we must impose that the parameters involved in the Dunkl derivatives must take the values $\mu_x\in {\mathbb{N}}$ and $\mu_y\in {\mathbb{N}}$ or $\mu_x\in \{\frac{1}{2},\frac{3}{2},...\}$ and $\mu_y\in\{\frac{1}{2},\frac{3}{2},...\}$.\\

The above results fulfill our purpose of obtaining the analytical solutions and the energy spectrum of the relativistic Dirac-Dunkl equations (\ref{eq1}) and (\ref{eq2}) for the case $\omega>\frac{\omega_c}{2}$.

\section{Exact Analytical Solutions for the cases $\omega=\omega_c/2$ and $\omega_c/2>\omega$}

In this Section we report our results for the cases $\omega=\frac{\omega_c}{2}$ and $\frac{\omega_c}{2}>\omega$, obtained in a similar way to those computed in Section 3. These results are reported in Table 1, where we have defined $G_\pm(\rho)\equiv e^{-\frac{1}{2}X}\rho^{A_\pm-\mu_+}$, $x=\frac{m\bar{\omega}\rho^2}{\hbar}\equiv \frac{m(\frac{\omega}{2}-\omega)\rho^2}{\hbar}$, $A_\pm=\sqrt{\lambda_\pm^2+\mu_\pm^2}$, $\mu_\pm=\mu_x\pm\mu_y$, $B_\pm=1+\frac{2\hbar\omega}{mc^2}A_\pm$.

\begin{table}[!htbp]
\begin{center}
\begin{tabular}{|c|c|c|c|c|}\hline
($s_x$, $s_y)$ & $\psi_{1,k}^{s_x,s_y}(\rho)$ & $\psi_{2,k'}^{s_x,s_y}(\rho)$ & $E_{k}^{s_x,s_y}$ & $E_{k}^{s_x,s_y}=E_{k'}^{s_x,s_y}\Rightarrow$\\ \hline
(1,1)&$G_+(\rho)L^{A_+}_k(x)$&$G_+(\rho)L^{A_+}_{k'}(x)$&$\pm mc^2\sqrt{B_++\frac{2\hbar\omega}{mc^2}(2k+\mu_+-\lambda_++2)}$&$k'=k+\mu_++1$\\ \hline
(-1,-1)&$G_+(\rho)L^{A_+}_k(x)$&$G_+(\rho)L^{A_+}_{k'}(x)$&$\pm mc^2\sqrt{B_++\frac{2\hbar\omega}{mc^2}(2k-\mu_+-\lambda_++2)}$&$k'=k-\mu_++1$\\ \hline
(1,-1) &$G_-(\rho)L^{A_-}_k(x)$&$G_-(\rho)L^{A_-}_{k'}(x)$&$\pm mc^2\sqrt{B_-+\frac{2\hbar\omega}{mc^2}(2k+\mu_--\lambda_-+2)}$&$k'=k+\mu_-+1$\\ \hline
(-1,1)&$G_-(\rho)L^{A_-}_k(x)$&$G_-(\rho)L^{A_-}_{k'}(x)$&$\pm mc^2\sqrt{B_-+\frac{2\hbar\omega}{mc^2}(2k-\mu_--\lambda_-+2)}$&$k'=k-\mu_-+1$\\ \hline
\end{tabular}
\caption{Shows the eigenfunctions and energy spectrum of the Dirac-Dunkl oscillator in (2+1)-dimensions for the case $\frac{\omega_c}{2}>\omega$.}
\end{center}
\end{table}

From equation (\ref{matriz}) we observe that when $\tilde{\omega}=0$, $H_D$ reduces to the Hamiltonian of a relativistic free particle. In fact, the solutions of  equations (\ref{eq1}) and (\ref{eq2}) with $\tilde{\omega}=0$ represent the states of a Dunkl-relativistic free particle of energy $E$. The solutions to these equations for $(s_x,s_y)=(1,1)$ and $(s_x,s_y)=(-1,-1)$ are
\begin{equation}
\psi_{1,2}^{s_x,s_y}(\rho)=\rho^{-\mu_+}J_{B_+}(\sqrt{2\tilde{\mathcal{E}}}\rho),
\end{equation}
and for $(s_x,s_y)=(1,-1)$ and $(s_x,s_y)=(-1,1)$
\begin{equation}
\psi_{1,2}^{s_x,s_y}(\rho)=\rho^{-\mu_+}J_{B_-}(\sqrt{2\tilde{\mathcal{E}}}\rho),
\end{equation}
with $J_{B_\pm}(\sqrt{2\tilde{\mathcal{E}}}\rho)$ the bessel $J$ functions and $\tilde{\mathcal{E}}=\frac{E^2-m^2c^4}{2\hbar^2 c^2}$.

We emphasize that the Dirac-Dunkl oscillator spinor components reported in the present Section and those reported in the previous one are finite at the coordinates origin and $\mathcal{L}^2$ integrable according to the scalar product defined in equation (\ref{sp}). Further, all the results obtained in Sections 3 and 4 agree with the energy spectrum and eigenfunctions of the relativistic oscillator coupled to an external magnetic field when we set $\mu_x=\mu_y=0$ \cite{Mandal,NOS3}.

\section{Non-Relativistic Limit of the Dirac-Dunkl oscillator}

The purpose of this Section is to analyze the non-relativistic limit of the Dirac-Dunkl oscillator. To do this,
we observe that equation (\ref{klein1}) can be written in the following form
\begin{equation}
\left(E^2-m^2c^4\right)\psi_1=2mc^2\left(\frac{P_D^2}{2m}+\frac{m\omega^2\rho^2}{2}-\tilde\omega L_z^D-\hbar\tilde\omega\mu_xR_x-\hbar\tilde\omega\mu_yR_y-\hbar\omega\right)\psi_1, \label{xaprox}
\end{equation}
where $L_z^D\equiv -\hbar \mathcal{J}$ and $P_D^2\equiv (-i\hbar \nabla_D)^2$. The non-relativistic limit is obtained by assuming that
$E=mc^2+\delta E$, with $\delta E\ll mc^2$. This implies that $E^2-m^2c^4\approx 2mc^2+O((\delta E)^2)$. Under these assumptions, equation (\ref{xaprox})
leads to
\begin{eqnarray}
\delta E\psi_1 &\approx \left(\frac{P_D^2}{2m}+\frac{m\omega^2\rho^2}{2}-\tilde\omega L_z^D-\hbar\tilde\omega\mu_xR_x-\hbar\tilde\omega\mu_yR_y-\hbar\omega\right)\psi_1 \\
&=\left(H_{dunkl}^{2D}-\tilde\omega L_z^D-\hbar\tilde\omega\mu_xR_x-\hbar\tilde\omega\mu_yR_y-\hbar\omega\right)\psi_1,
\end{eqnarray}
where we have introduced the non-relativistic definition of the two-dimensional Dunkl oscillator \cite{GEN1}.
If we reintroduce the total energy $\delta E=E-mc^2$, we obtain the effective Hamiltonian in the non-relativistic limit for the upper spinor component
\begin{equation}
H_{eff}^{(\uparrow)}=mc^2+\left(H_{dunkl}^{2D}-\tilde\omega L_z^D-\hbar\tilde\omega\mu_xR_x-\hbar\tilde\omega\mu_yR_y-\hbar\omega\right).
\end{equation}
Similarly we proceed to obtain the non-relativistic limit for the lower component $\psi_2$. This is obtained by setting $E=-mc^2+\delta E$, where $\delta E\ll mc^2$. Thus, $E^2-m^2c^4\approx-2mc^2+O((\delta E)^2)$. By substituting these results in equation (\ref{klein2}), we can show that
the effective Hamiltonian for the lower spinor component in the non-relativistic regime is
\begin{equation}
H_{eff}^{(\downarrow)}=-mc^2-\left(H_{dunkl}^{2D}-\tilde\omega L_z^D+\hbar\tilde\omega\mu_xR_x+\hbar\tilde\omega\mu_yR_y+\hbar\omega\right).
\end{equation}

As it has been shown, our Hamiltonian can be properly reduced to the non-relativistic Hamiltonian of the harmonic oscillator in two dimensions. This reduction includes, as it is usual, an angular term, which for this case is the Dunkl angular momentum in the $z$-axis. Moreover, the non-relativistic limit of the Dirac-Dunkl oscillator contains terms related with the reflection operators due to the Dunkl derivatives.

\section{Concluding Remarks}

In the present paper we generalized the problem of the Dirac oscillator coupled to a uniform magnetic field by changing the standard partial derivatives by the Dunkl derivatives. From some properties of the reflection operators, we were able to decouple the differential equations for each spinor component. Due to the characteristics of our problem we divided your study into two cases, $R_x=R_y$ and $R_x=-R_y$. By using known results on the eigenvalues and eigenfunctions of the angular operator $\mathcal{J}=i(xD_y-yD_x)$, we obtained in a closed form the eigenfunctions and energy spectrum for all of the cases of the relativistic Dirac-Dunkl oscillator. In this way we are introducing a new exactly solvable quantum mechanical problem. We emphasize that, due to the generality of the Hamiltonian that we have studied in the present paper, it is properly reduced to several particular problems. Thus, if we put off the magnetic field, we obtain the generalized two-dimensional Dirac-Dunkl oscillator. Now, if we consider a fermion in a uniform magnetic field (without Dirac oscillator), we obtain the Dirac-Dunkl Landau levels. Also, we notice that when we set $\mu_x=\mu_y=0$ in our Dirac-Dunkl Hamiltonian, our results reduce to the spectrum and the eigenfunctions of a relativistic oscillator coupled to an external magnetic field \cite{Mandal,NOS3}.

In Ref. \cite{Delgado} the relativistic problem of the Dirac oscillator plus a constant magnetic field has been solved. In that work, the authors showed that the Dirac oscillator contributes with a left-handed chirality (anti-Jaynes-Cummings) and the magnetics field contributes with a right-handed chirality (Jaynes-Cummings). Thus, there exists an interplay of opposite chirality which is used to study a relativistic phase transition. The play between Jaynes-Cummings and anti-Jaynes-Cummings is determined by the difference between $\omega$ and $\omega_c/2$, being the critical regime when
$\omega=\omega_c/2$. In this case, the two-dimensional fermion behaves as a free relativistic particle. It has been shown that angular momentum in the $z$-direction has the role of a parameter which witnesses the quantum phase transition by changing its sign. The Dirac-Dunkl oscillator studied in the present paper is a generalization of the Hamiltonian considered in Ref. \cite{Delgado}. Also, we solved our problem for all the cases $\omega>\omega_c/2$, $\omega<\omega_c/2$ and $\omega=\omega_c$. We have shown that the Dirac-Dunkl Hamiltonian has the critical regime when $\omega=\omega_c/2$. In fact, it can be shown that our Hamiltonian can be written as a competition of generalized Jaynes-Cummings and anti-Jaynes-Cummings terms and that the Dunkl angular momentum $\mathcal{J}=i(xD_y-yD_x)$ is the parameter which witnesses the quantum phase transition. These points will be reported in a future work.

Finally, we point out that the Coulomb problem \cite{GEN4} and the problem of a relativistic particle coupled to a uniform magnetic field plus a Coulomb potential \cite{tai1,tai2} in $(2+1)$ dimensions can be studied by the idea introduced in the present paper. These are works in progress.

\section{Appendix. The Dunkl angular momentum operator $\mathcal{J}$ and its eigenfunctions}

In this Appendix we give the main results on the eigenvalues and eigenfunctions of the Dunkl angular operator $\mathcal{J}\equiv i(xD_y-yD_x)$ reported in Ref. \cite{GEN1}. In polar coordinates it takes the form
\begin{equation}
\mathcal{J}=i(\partial_\phi+\mu_y\cot\phi(1-R_y)-\mu_x\tan\phi(1-R_x)).
\end{equation}
It can be shown that the square of this operator can be written as
\begin{equation}
\mathcal{J}^2=2B_\phi+2\mu_x\mu_y(1-R_xR_y),
\end{equation}
where
\begin{equation}
B_\phi=-\frac{1}{2}\partial_\phi^2+\left(\mu_x\tan\phi-\mu_y\cot\phi\right)\partial_\phi+\frac{\mu_x(1-R_x)}{2\cos^2\phi}+\frac{\mu_y(1-R_y)}{2\sin^2\phi}.\label{be}
\end{equation}
Since the operator $R_xR_y$ commutes with the operator $\mathcal{J}$, its eigenvalues and eigenvectors are search in the form
\begin{equation}
\mathcal{J}F_\epsilon=\lambda_\epsilon F_\epsilon,\label{angular}
\end{equation}
being $\epsilon\equiv s_xs_y=\pm 1$, and $s_x$, $s_y$ the eigenvalues of the reflection operators $R_x$ and $R_y$, respectively.
\begin{itemize}
\item If $R_x=R_y$, then $\epsilon=1$. In this case, the solutions of equation (\ref{angular}) are given by
\begin{eqnarray}
F_+=\Phi_n^{++}(\phi)\pm i\Phi_n^{--}(\phi),\hspace{.5ex}\\
\lambda_+=\pm2\sqrt{n(n+\mu_x+\mu_y)}
\end{eqnarray}
where $n\in {\mathbb{N}}$. The functions $\Phi_n^{++}$ and $\Phi_n^{--}$ are given by
\begin{eqnarray}
\Phi_n^{++}(x)=\sqrt{\frac{(2n+\mu_x+\mu_y)\Gamma{(n+\mu_x+\mu_y)n!}}{2\Gamma{(n+\mu_x+1/2)}\Gamma{(n+\mu_y+1/2)}}}P_n^{(\mu_x-1/2,\mu_y-1/2)}(x),\hspace{18ex}\label{masmas}\\
\Phi_n^{--}(x)=\sqrt{\frac{(2n+\mu_x+\mu_y)\Gamma{(n+\mu_x+\mu_y+1)(n-1)!}}{2\Gamma{(n+\mu_x+1/2)}\Gamma{(n+\mu_y+1/2)}}}\sin\phi\cos\phi P_{n-1}^{(\mu_x+1/2,\mu_y+1/2)}(x),\label{menmen}
\end{eqnarray}
where $P_n^{(\alpha,\beta)} (x)$ are the classical Jacobi polynomials and $x=-\cos2\phi$. It is understand that $P_{-1}^{(\alpha,\beta)}(x)=0$ and hence that $\Phi_0^{--}=0$
\item
For $R_x=-R_y$, $\epsilon=-1$. In this case it has been shown that
\begin{eqnarray}
F_-=\Phi_n^{-+}(\phi)\mp i\Phi_n^{+-}(\phi),\hspace{3ex}\\
\lambda_-=\pm2\sqrt{(n+\mu_x)(n+\mu_y)},
\end{eqnarray}
where $n\in \{\frac{1}{2},\frac{3}{2},...\}$. The eigenfunctions $\Phi_n^{-+}$ and $\Phi_n^{+-}$ are given by
\begin{eqnarray}
\Phi_n^{-+}(x)=\sqrt{\frac{(2n+\mu_x+\mu_y)\Gamma{(n+\mu_x+\mu_y+1/2)(n-1/2)!}}{2\Gamma{(n+\mu_x+1)}\Gamma{(n+\mu_y)}}}\cos\phi P_{n-1/2}^{(\mu_x+1/2,\mu_y-1/2)}(x),\label{menmas}\\
\Phi_n^{+-}(x)=\sqrt{\frac{(2n+\mu_x+\mu_y)\Gamma{(n+\mu_x+\mu_y+1/2)(n-1/2)!}}{2\Gamma{(n+\mu_x)}\Gamma{(n+\mu_y+1)}}}\sin\phi P_{n-1/2}^{(\mu_x-1/2,\mu_y+1/2)}(x).\hspace{.5ex}\label{masmen}
\end{eqnarray}
\end{itemize}

\end{document}